\documentclass[showpacs, preprintnumbers, nofootinbib]{revtex4}
\usepackage{amsmath,amssymb}
\usepackage{graphicx}
\usepackage{bm}
\usepackage{natbib}
\newcommand{\beq}{\begin{equation}}
\newcommand{\eeq}{\end{equation}}
\newcommand{\beqa}{\begin{eqnarray}}
\newcommand{\eeqa}{\end{eqnarray}}

\def\fun#1#2{\lower3.6pt\vbox{\baselineskip0pt\lineskip.9pt
  \ialign{$\mathsurround=0pt#1\hfil##\hfil$\crcr#2\crcr\sim\crcr}}}

\begin{document} 
\title{
Optimal Extraction of Cosmological Information from Supernova Data in the Presence of Calibration Uncertainties} 
\author{Alex G. Kim and Ramon Miquel} 
\affiliation{Physics Division, Lawrence Berkeley National Laboratory, 1 Cyclotron Rd.,
Berkeley, CA 94720} 
\email{AGKim, RMiquel@lbl.gov}

\date{\today}

\begin{abstract} 
\noindent
We present a new technique to extract the cosmological information from high-redshift supernova data in the
presence of calibration errors and extinction due to dust. While in the traditional technique the distance modulus of
each supernova is determined separately, in our approach we determine all distance moduli at once, in a process that achieves
a significant degree of self-calibration. The result is a much reduced sensitivity of the cosmological parameters to the
calibration uncertainties. As an example, for a strawman mission similar to that outlined in the SNAP satellite proposal, 
the increased precision obtained with the new approach is roughly equivalent to a factor of five decrease in the
calibration uncertainty.
\end{abstract} 

\pacs{98.80.Es, 97.60.Bw, 95.75.Pq}     

\maketitle 
\section{Introduction} \label{sec:intro}
The study of type-Ia supernovae provided the first indication that the
expansion of the Universe is
accelerating~\cite{42SNe_98,Riess_acc_98}. This conclusion is
supported by the combination of Cosmic Microwave Background
results~\cite{Bennett_etal:2003} and large-scale structure
measurements~\cite{Tegmark:2004}.  Results from more recent supernova
surveys~\cite{Knop_etal:2003,Riess_and_others:2004} further strengthen
the evidence for the accelerated expansion of the Universe and the
necessity for some mysterious mechanism that drives it, code-named
``dark energy.''

Supernova magnitudes at peak brightness and redshifts are the
ingredients that allow their use as cosmological probes. 
Peak magnitudes can be corrected with the ``stretch''~\cite{42SNe_98},
$\Delta m_{15}$~\cite{Phillips99}, or MLCS~\cite{rpk96} techniques to turn
type-Ia supernovae into standardized candles to a level around
0.10--0.15 mag (10--15\% in flux).  Therefore, measuring the supernova
apparent magnitude provides the distance and hence the lookback time
to when the explosion occurred, while measuring its redshift $z$
provides the scale factor $a = (1+z)^{-1}$ at the time the supernova
light was emitted.  These two statistics for each supernova from a set
that spans a range of redshifts provide a mapping of the cosmic
expansion with time.

Neglecting for the moment any secondary effects, 
one can relate the apparent peak magnitude $m(z)$
of a supernova at redshift $z$ to the luminosity distance $d(z) =
r(z)(1+z)$, where $r(z)$ is the dimensionless comoving distance, through:
\begin{equation}
m(z) \equiv -2.5 \log_{10} {\left({\cal F}/{\cal F}_0\right)} 
     = {\cal M} + \mu(z)            
     = {\cal M} + 5\log_{10}d(z) + 25\,   
\label{eq:m}
\end{equation}
where $\cal F$ is the observed supernova flux and ${\cal F}_0$ is the flux
corresponding to zero magnitude of the magnitude system. ${\cal M}
\equiv M -
5\log_{10}\left[H_0/100\right.$km~s$^{-1}$Mpc$\left.^{-1}\right]$,
can be regarded as an unknown nuisance constant, 
$M$ being the absolute supernova magnitude, 
while $\mu$ is the dimensionless
supernova distance modulus.
With this definition $r(z)$ is also
dimensionless and does not depend on $H_0$:
\beqa
\label{eq:r}
r(z)     &=& \int_0^z dz' H_0 \frac{a}{\dot{a}} = \int_0^z dz' \frac{H_0}{H(z')}\ , 
\\
\left(H(z)/H_0\right)^2 &=& \Omega_m \, (1+z)^3 
         + \left(1-\Omega_m\right) \, \exp\left[3\int_0^z \frac{dz'}{1+z'} \left(1+w(z')\right)\right] , 
\nonumber
\eeqa
where we have assumed a flat universe and $w(z)$ is the equation of state of the dark energy, 
$w(z)=-1$ for a cosmological constant. In more general models 
$w(z)$ would simply be a function that captures the effect of dark energy.
Following~\cite{Polarski:2001,Linder:2003} we consider the
parameterization
\beq
w(z) = w_0 + w_a \left(1-a(z)\right) \ ,
\label{eq:w}
\eeq
where $w_0$ and $w_a$ are constants that determine, respectively, 
the dark-energy equation of state now and its evolution.

In a real observation there are many more effects that have to be
accounted for: weak lensing, Milky Way extinction, etc. In this paper,
we will concentrate on two (inter-related) effects: host-galaxy
extinction due to dust and mis-calibration.  Some of the other effects
are treated in~\cite{Kim_etal:03}.  Dust in the supernova host galaxy will
dim the light coming from the supernova, therefore biasing its
measured magnitude toward larger values. However, ordinary dust dims
light of different wavelengths differently, absorbing more blue light
than red light. By measuring the supernova flux in several broadband
filters, one can determine the reddening and hence the overall dimming
that dust creates. In order to proceed with the last step two things
are needed: a good understanding of the gross features of
non-extincted type-Ia supernova spectra, and a knowledge of the
relationship between differential dimming across wavelengths and
absolute extinction. We will assume a perfect knowledge of the former
and for the latter we will use the standard Cardelli, Clayton and
Mathis model~\cite{card89}.  In this model, the extinction $A$ (that
is, the dimming in magnitude units) at a wavelength $\lambda$ can be
written as:
\beq
A(\lambda) = A_V \cdot \left(a(\lambda)+b(\lambda)/R_V\right)
\eeq
where the parameters $A_V$ and $R_V$ control, respectively, the
overall amount of dimming and its dependence on wavelength, and
$a(\lambda)$ and $b(\lambda)$ are known functions.  The value of $A_V$
will be different for different supernovae depending on the 
properties of the host
galaxy and the position of the supernova inside it. The value of $R_V$
will depend on the type of dust (grain size, for instance), and
observations show~\cite{Falco:1999} that it varies significantly
across redshift and between galaxies.  In the following, we will
assume no prior knowledge on either $A_V$ or $R_V$. Then, given enough
independent measurements of the magnitudes of a given supernova $i$ in
several broadband filters centered around wavelengths $\lambda_\alpha$, one
can in principle determine directly from the observations the dust
parameters $A_V^i$, and $R_V^i$ for supernova $i$, together with its
distance modulus.

Filter calibration uncertainties affect the measured magnitude in
each passband. Therefore, they affect not only the overall peak
magnitude but also the extinction correction that, as mentioned
above, depends on understanding the supernova colors (defined as the
ratio of the supernova flux in two different filters).  The
calibration error in a given observer filter will affect each
supernova in a different (but correlated) way, since a given passband
in the supernova rest frame will be observed with a different filter in 
the observatory depending on the supernova redshift.
Calibration uncertainties are going to be among the limiting
uncertainties~\cite{Kim_etal:03} in the new generation of supernova surveys being
currently planned.

In the standard way of extracting cosmological information from
supernova data, the data for a given supernova $i$ is used to
determine the values of $\mu^i$, $A_V^i$, and $R_V^i$ and their
uncertainties. Then the calibration error is propagated through that
supernova's measurements.  As a result, a non-diagonal covariance
matrix for the set of distance moduli $\{\mu^i\}$ is obtained. The
measurements of all the $\{\mu^i\}$ and their covariance matrix are
then used in a fit to determine the cosmological parameters
$\Omega_m$, $w_0$, and $w_a$, using Eqs.~(\ref{eq:m}), (\ref{eq:r})
and (\ref{eq:w}).

In our approach, all $\{\mu^i\}$ are determined simultaneously. In
this way, all information available is used in an optimal way, which
achieves a certain degree of self-calibration.  The cosmological
parameter fit then follows as above.  Equivalently, one can just
bypass completely the step of extracting the distance moduli and
determine directly the cosmological parameters from the set of
magnitudes measured for all supernovae $\{m^i_j\}$ ($i$ runs over supernovae,
whereas $j$ runs over passbands in the supernova rest frame).  In
Section~\ref{sec:cali} we present a fiducial supernova mission: the
dataset it provides and the algorithms through which the dataset is
used to measure the dark energy.  Within that framework, we elaborate
further on the calibration error and its interplay with extinction.
Section~\ref{sec:anal} describes both the traditional data analysis
method and our new approach.  Results using both approaches are given
in Section~\ref{sec:results} for a fiducial mission and the
statistical advantage that the new, optimal method affords becomes
apparent. Finally, we draw our conclusions in Section~\ref{sec:summ}.
%
\section{Calibration and supernova cosmology} \label{sec:cali}
Although our results have a more general validity, in the remainder we
discuss a specific strawman mission along the lines of the SNAP
satellite proposal~\cite{omnibus:2004}.  The mission observes 2000 type-Ia
supernovae over a redshift range from 0 up to 1.7 with a set of nine
logarithmically-spaced broadband filters centered at wavelengths
$\lambda_\alpha = \lambda_0 \cdot (1+0.16)^\alpha$, $\alpha=0,\ldots,8$, with
$\lambda_0 = 440$~nm corresponding to the $B$ band.  For every
supernova, the analysis is based on the $B$ passband in the
supernova rest frame and any other band redder than this which, after
redshifting to the observer frame, still ends up overlapping one of
the nine filters. So only optical and infrared information in the
supernova rest frame is used. The current lack of understanding
of supernova heterogeneity in the UV prevents the robust usage of
shorter wavelengths (though see \cite{Jha(2002),Guy:2005} for perspectives for
using UV light curves to aid in measuring supernova distances).

Therefore, $9-k$ filters are used for a supernova with $z$ between
$z_k$ and $z_{k+1}$, where $z_k = (1+0.16)^k-1$, $k=0,\dots,6$.  With
this choice of filters, at least three bands are available for each
supernova.  The calibration uncertainty enters through the zero points
of the filters, ${\cal Z}_\alpha = 2.5\log_{10} {\cal F}_{\alpha 0}$, 
$\alpha=0,\ldots,8$, where ${\cal F}_{\alpha 0}$ 
is the observed 
flux measured in filter $\alpha$ for a standard object of known magnitude 0.
So in order to specify a calibration error, one has to specify the
$9\times 9$ covariance matrix for $\{\cal{Z}_\alpha\}$.

We are now in a position to write the equation that will relate
the measured light-curve-shape-corrected~\cite{42SNe_98,Phillips99,rpk96}
and K-corrected~\cite{Nugent:2002} peak
magnitudes $m^i_j$ with the parameters of our model
(details on K-correction uncertainties, which we will not consider here,
can be found in~\cite{Davis:2005}):
\beq
\label{eq:master}
m^i_j + {\cal Z}_{\alpha(i,j)} = {\cal M}_j + \mu^i(z^i;\Omega_m,w_0,w_a) 
                               + A_V^i \cdot a(\lambda_j) + B_V^i \cdot b(\lambda_j)\ .
\eeq
Indices $i$ and $j$ run respectively over supernovae and
passbands. ${\cal M}_j$ is $\cal M$ for passband $j$: ${\cal M}_j
\equiv M_j  -
5\log_{10}\left[H_0/100\right.$km~s$^{-1}$Mpc$\left.^{-1}\right]$,
where $M_j$ is the absolute magnitude of a type-Ia supernova measured
in passband~$j$. $B_V$ is defined as $A_V/R_V$ and makes the problem
linear on the dust parameters. Finally, $\alpha(i,j)$ is the index of
the filter in which, after K-correction, rest frame passband $j$ of supernova 
$i$ is measured.
For $z_k \leq z^i < z_{k+1}$, we have $\alpha(i,j) = j+k$, and the index $j$ 
runs from 0 to $9-k-1$, so that there are $9-k$ measurements for supernova $i$.

The left-hand side of Eq.~(\ref{eq:master}) contains the
measurements,
while in the right-hand side we have the parameters to be determined:
${\cal M}_j$, $j=0,\ldots,8$, and $\mu^i$, $A_V^i$, $B_V^i$ for each
supernova.  Alternatively, the fit parameters $\mu^i$ can
be replaced with the fit parameters $\Omega_m$, $w_0$, and $w_a$
with $\mu^i \to \mu(z^i;\Omega_m,w_0,w_a)$ where the latter
is the predicted magnitude as a function of the cosmological parameters.
A careful examination of Eq.~(\ref{eq:master}) reveals
that there are two non-independent parameters. Specifically, a
simultaneous shift in all $A_V^i$ to $A_V^i + \delta$, can be
compensated by a shift of all ${\cal M}_j$ to ${\cal M}_j - \delta\cdot
a(\lambda_j)$. Analogously, a change $B_V^i \to B_V^i + \eta$ is
exactly compensated by the shift ${\cal M}_j \to {\cal M}_j -
\eta\cdot b(\lambda_j)$. Therefore, one has to eliminate two
parameters. To do so, we assume that for a nearby supernova we know 
that it has no dust (for instance because it lies in an
elliptical galaxy), or equivalently that its dust extinction is
perfectly known.  For our analysis, we choose to assume that for the
supernova with index $i=0$ it holds $A_V^0 = B_V^0 = 0$.  Furthermore,
for simplicity, we assume that we know the peak magnitudes for that
supernova, $m^0_j$, perfectly well. Results do not depend strongly on
this assumption, which can be justified because the statistical errors
in the photometry of a nearby supernova are going to be much smaller
than those for the high-redshift ones.  We can now subtract
from Eq.~(\ref{eq:master}) the particular equation for $i=0$, to get
\beq
\label{eq:master2}
m^i_j - m^0_j + {\cal Z}_{\alpha(i,j)} - {\cal Z}_j = \mu^i(z^i;\Omega_m,w_0,w_a) - \mu^0 
                               + A_V^i \cdot a(\lambda_j) + B_V^i \cdot b(\lambda_j)\, ,
\eeq
where we have taken into account ${\cal Z}_{\alpha(0,j)} = {\cal Z}_j$
when supernova 0 has a low redshift $z^0 < z_1=0.16$. While the
$m^0_j$ magnitudes are assumed known, $\mu^0$ is not and it becomes an
additional free parameter.  Now the index $i$ runs from one to the
number of supernovae minus one. The number of free parameters has been
decreased by two with respect to Eq.~(\ref{eq:master}).  Written in
this form, it also becomes clear that only eight color
calibrations (${\cal Z}_\alpha - {\cal Z}_\beta$) rather than nine absolute 
calibrations are necessary for the
cosmology determination.  This is the framework that we use in the
following.
\section{Simulated data analysis} \label{sec:anal}
\subsection{Supernova-by-supernova procedure} \label{sec:trad}
In standard supernova-cosmology analysis, the distance to each
supernova is determined independently.  In our implementation of the
traditional analysis procedure, values for $A_V^i$, $B_V^i$, and
$\mu^i - \mu^0$ are determined from the set of magnitudes $\{m^i_j\}$
for supernova $i$.  Only statistical errors $\sigma^i_{j,\,{\mathrm {stat}}}$ 
in the measurements of the
magnitudes are included in the fit.  A Monte Carlo process is then
used to compute the additional covariance matrix for the set of
$\{\mu^i\}$ due to the zero-point 
calibration uncertainties $\sigma_{\alpha,\,{\mathrm {cal}}}$,
where the index $\alpha$ labels the filters. 
Finally, the cosmology
fit is performed on all $\{\mu^i\}$ using the covariance matrix that
includes calibration errors plus an additional intrinsic dispersion
$\sigma^i_{\mathrm {int}}$
for each supernova distance modulus that takes into account known
supernova-to-supernova
variability. The free parameters in the cosmology fit are
$\mu^0$, $\Omega_m$, $w_0$, and $w_a$. For the fit we use the standard
package {\tt minuit}~\footnote{{\tt http://wwwinfo.cern.ch/asdoc/minuit/minmain.html}}. 
We have checked that a Fisher
matrix calculation provides compatible error estimates.
\subsection{Simultaneous procedure} \label{sec:new}
In the new approach, all supernova data are used at once in order to
determine all $A_V^i$, $B_V^i$, and $\mu^i - \mu^0$. The input
covariance matrix for the $m^i_j$ includes statistical errors (which
are diagonal), calibration error and intrinsic dispersion.  For a
large simulated data set of about 2000 supernovae, this treatment is
rather impractical, since the input covariance matrix is about
12000$\times$12000 and non-diagonal, which makes it difficult to store and invert
efficiently. Instead we can treat both the intrinsic dispersion and
the calibration error as new free fit parameters, constrained within
the known uncertainties. Then one can build a $\chi^2$ function whose
covariance matrix is diagonal:
\beqa
\label{eq:chi2}
\chi^2 &=&  \sum_{ij}\left(\frac{m^i_j - m^0_j -
                        \overline{m}^i_j\left(\mu^i-\mu^0, A_V^i, B_V^i, S^i, {\cal Z}_\alpha\right)}
                               {\sigma^i_{j,\,{\mathrm {stat}}}} \right)^2 
       + \sum_i\left(\frac{S^i}{\sigma^i_{\mathrm {int}}}\right)^2 + 
       \sum_\alpha \left(\frac{{\cal Z}_\alpha}{\sigma_{\alpha,\,{\mathrm {cal}}}}\right)^2 \ , \\
\overline{m}^i_j  &=& \mu^i\left(z^i;\Omega_m,w_0,w_a\right) - \mu^0
             + A_V^i \cdot a\left(\lambda_j\right) + B_V^i \cdot b\left(\lambda_j\right) 
		+ S^i - {\cal Z}_{\alpha(i,j)} + {\cal Z}_j \ ,
\label{eq:master3}
\eeqa
where in Eq.~(\ref{eq:chi2}), again the indices $i$ and $j$ run over
supernovae and passbands respectively, $\alpha$ runs over filters, and
$S^i$ is the new free parameter that accounts for the intrinsic
supernova dispersion, one new parameter per supernova.  Now the
$\{{\cal Z}_\alpha\}$ are considered as nine additional free
parameters, constrained to vary within the calibration precision.
The uncertainties $\sigma^i_{j,\,{\mathrm {stat}}}$, $\sigma^i_{\mathrm {int}}$, and
$\sigma_{\alpha,\,{\mathrm {cal}}}$ are as defined in Section~\ref{sec:trad}.
Typically, for 2000 supernovae,
we now have about 12000 measurements and over 8000 unknowns. After all
$\{\mu^i-\mu^0\}$ are determined, the cosmology fit proceeds as in
Section~\ref{sec:trad}, but now using the $\{\mu^i\}$ covariance matrix
from the overall fit to all distance moduli.

Alternatively, and equivalently, one can do without the determination of the distance moduli, and just fit all
the magnitudes directly to the cosmological parameters and the other nuisance parameters, by using again the $\chi^2$ 
in Eqs.~(\ref{eq:chi2},\ref{eq:master3}). Now the free parameters in the fit are all the $A_V^i$, $B_V^i$, $S^i$
and ${\cal Z}_\alpha$,
as well as $\mu^0$, $\Omega_m$, $w_0$, and $w_a$. This is the final approach that we have used. A Fisher matrix
calculation following these lines runs on a 64-bit\footnote{Use of 128-bit (double) precision arithmetic is crucial 
in the process of inverting the about 6000$\times$6000 heavily-correlated Fisher matrix.},
3~GHz Intel Xeon processor in about one hour for 2000 supernovae.
%
%
\section{Results} \label{sec:results}
\subsection{Calibration models} \label{sec:models}
In this section we are going to present results based on two extreme
calibration models. We will not attempt to construct a realistic
model from observations of standard stars or calibrators. Instead,
we will specify the calibration model by its 9$\times$9 zero-point
covariance matrix.

In the first model, we assume that the zero-point covariance matrix is
fully diagonal. While this is surely an unrealistic scenario (the same
fundamental calibrator is probably going to be used for more than one
filter, if not all), there will still be a diagonal component in the
full calibration covariance matrix, originating for instance from
Poisson uncertainties in the measurement of the fundamental
calibrator.  For simplicity, we will take the calibration errors in
all filters to be equal, so that the zero-point covariance matrix will
be simply $V_{\alpha\beta} =
\sigma^2_{\mathrm{cal}}\cdot \delta_{\alpha\beta}$.  We will call this
model the ``Diagonal'' model.

The other model we consider goes to the opposite extreme, and assumes
that all sources of error during the calibration process come from the
uncertainty in a single parameter, such as the temperature of a hot
white dwarf used as a calibrator. In this case the zero-point
parameters are replaced with a single temperature parameter $T$ and
Eq.~(\ref{eq:chi2}) has to be modified to replace the last term with
a term $(T-T_0)^2/\sigma_T^2$ where $T_0$ is the given temperature and
$\sigma_T$ the precision with which it is known. Then, in
Eq.~(\ref{eq:master3}) the zero points ${\cal Z}_\alpha$ are turned into
functions of the temperature parameter by integrating in the
corresponding filter  
the black body spectrum of the white dwarf
with that temperature.  We will call this model the ``Temperature''
model.

As mentioned above, these are two very extreme toy models for the
calibration covariance matrix.  The true covariance matrix will be
much more complicated and will depend on the specific details of the
mission and on the set of standard calibrators used. However, these
two models should span the range of possible calibration error models.
\subsection{No calibration error} \label{sec:nocal}
To start with, we consider the case with no calibration error. That
is, we fix all parameters $\{{\cal Z}_\alpha\}$ in
Eq.~(\ref{eq:chi2}) or equivalently, we take the limit
$\sigma_{\alpha ,\, \mathrm{cal}} \to 0$. Let us now specify
completely our fiducial survey. Table~\ref{tab:z} gives the number of
supernovae in each redshift bin, based on the numbers in~\cite{Kim_etal:03}.
\begin{table}[t]
\caption[]{The redshift distribution $N(z)$ of the supernovae employed
in the fiducial survey.
The redshifts $z$ given in the table
correspond to the value for all supernovae in that bin.}
\label{tab:z}
\begin{center}
\begin{tabular}{|c||r|r|r|r|r|r|r|}
\hline
$z$    & 0.05 & 0.17  & 0.35    & 0.57  & 0.82  & 1.11  & 1.50 \\
\hline
$N(z)$ & 317  & 82    & 219     & 412   & 441   & 427   & 400  \\
\hline
\end{tabular}
\end{center}
\end{table}
All supernovae in a bin have been assigned the same actual redshift,
which is also reported in the table. In total there are 2298
supernovae, 300 of which would be obtained at low redshift by
ground-based observations like those of the Supernova Factory~\cite{Wood-Vasey:2004}. 
We consider that the calibration error in the
observation of these 300 additional supernovae is completely
correlated with the calibration error in the original 2000
supernovae. This is what will happen if the dominant calibration
uncertainty comes from the limited knowledge of the intrinsic
properties of the calibrators, and the same calibrators are used
in both surveys.

For the statistical errors in the magnitudes we choose $\sigma^i_{j,\,
\mathrm{stat}} = 0.01$ for all passbands in all supernovae.
This uncertainty encompasses uncorrelated statistical uncertainties
introduced from the photometry process (e.g.\ flatfield errors,
Poisson noise) and intrinsic uncorrelated supernova color dispersion.
The intrinsic dispersion in the supernova magnitude is
$\sigma^i_{\mathrm{int}} = 0.15$ for all supernovae, independent from
supernova to supernova and fully correlated for all passbands within
one supernova.  Other than the supernova measurements, we will include
in our synthetic data sample a measurement of the distance to the
surface of last scattering from the future Planck survey, with a
relative precision of $0.7\%$. Finally, we take as our fiducial cosmology
the flat $\Lambda$CDM model, $w_0=-1$, $w_a=0$, with $\Omega_m = 0.28$. 

The resulting uncertainties in the cosmological parameters $w_0$ and $w_a$ are given in the first two rows of
Table~\ref{tab:result}. 
\begin{table}[t]
\caption[]{Uncertainties in the cosmological parameter determination in the two analysis methods
for several values of the
calibration uncertainty in the ``Diagonal'' error model. }
\label{tab:result}
\begin{center}
\begin{tabular}{|c|c||c|c|}
\hline
$\sigma_{\mathrm{cal}}$ && SN by SN & Simultaneous \\
\hline
\hline
0.000 & $\sigma(w_0)$ & 0.064 & 0.064 \\
      & $\sigma(w_a)$ &  0.30 & 0.30  \\
\hline
0.001 & $\sigma(w_0)$ & 0.082 & 0.068 \\
      & $\sigma(w_a)$ & 0.40  & 0.33 \\
\hline
0.005 & $\sigma(w_0)$ & 0.099 & 0.071 \\
      & $\sigma(w_a)$ & 0.59  & 0.43  \\
\hline
0.010 & $\sigma(w_0)$ & 0.11  & 0.075 \\
      & $\sigma(w_a)$ & 0.81  & 0.53  \\
\hline
\end{tabular}
\end{center}
\end{table}
When there are no calibration errors the results obtained with the two analysis methods fully
agree, as they should, given that in this case there is no information shared between any two supernovae,
and therefore it should not make a difference whether they are analyzed separately or all at once.
\subsection{Diagonal calibration model} \label{sec:diag}
For the ``Diagonal'' calibration error model, we will choose
$V_{\alpha\beta} = \sigma^2_{\mathrm{cal}}\cdot \delta_{\alpha\beta}$,
with $\sigma_{\mathrm{cal}}$ varying between 0.001 and 0.01. The
results again are in Table~\ref{tab:result}. Since now, because of the
calibration error, there is correlated information shared among the
supernovae, there is a clear advantage in analyzing all supernovae at
once, to the point that, for instance, the effect of a 0.005
calibration uncertainty using the simultaneous fit is reduced to
approximately the effect of a 0.001 calibration uncertainty when
considering each supernova individually. 

Inspection of the final
uncertainties
in ${\cal Z}_\alpha$ after the Fisher matrix calculation reveals the amount
of self-calibration achieved: for example when 0.010 is assumed as the prior
error in each ${\cal Z}_\alpha$, the final (posterior) ${\cal Z}_\alpha$ 
errors are reduced
to values ranging between 0.004 and 0.008, with large positive correlations
between neighboring filters.
\subsection{One-parameter calibration model} \label{sec:temp}
In the ``Temperature'' model, one single parameter, in this case the
temperature of a single $T_0=20000$K
hot white dwarf, defines the whole calibration
error matrix. In this case, the final precision in the cosmological
parameters is rather insensitive to the calibration error
as was found in \cite{Kim_etal:03}. For
instance, assuming a temperature error such that it results in an
uncertainty for the zero point of the first filter of
$\sigma_{0,\,\mathrm{cal}} = 0.01$ (and totally correlated
uncertainties in the other zeropoints) results in an inappreciable
change in the error on $w_0$ and $w_a$. Even if we assume
$\sigma_{0,\,\mathrm{cal}} = 0.10$, the uncertainties increase only
marginally from 0.064 to 0.066 ($w_0$) and from 0.30 to 0.31
($w_a$). These results hold when performing the analysis in either of
the two methods. The reason is that with one single parameter defining
the calibration error, self-calibration can occur already within a
single supernova, provided there are at least four passbands
available, which is the case for over 80\% of our supernovae.
\section{Conclusions} \label{sec:summ}
In summary, we have presented a new method of analysis of
high-redshift supernova data in the presence of calibration
uncertainties and dust extinction that performs the statistical
analysis of all data at once in order to self-calibrate the zeropoints
of the filter set. For a fiducial mission inspired on that specified
in the SNAP proposal, a significant reduction of the sensitivity of
the cosmological parameters to the calibration uncertainty is achieved
with the new method, equivalent to a reduction of about a factor five
in calibration uncertainty.  The advantage of the simultaneous
analysis is not only applicable to calibration uncertainties, but to
all sources of uncertainty that are non-trivially correlated between
supernovae.

Calibration
uncertainties are no longer treated as irreducible; the supernova data
provide an avenue for self-calibration of magnitude 
zeropoints. Assuming that type-Ia supernovae are standardizable candles 
in our passbands, their colors as a function of redshift can be predicted. 
Measurements of supernova colors at different redshifts therefore 
constrain the differences in 
zeropoints, $\{{\cal Z}_\alpha-{\cal Z}_\beta\}$. Some of the color 
information is spent in correcting for dust extinction of individual 
supernovae, however given enough supernovae observed in enough passbands, 
a significant improvement in calibration is still achieved.

This paradigm for simultaneous supernova analysis gives rise to new
questions that can be addressed in future studies. 
The interplay between the supernova sample, input calibration uncertainty,
cosmology priors, photometry uncertainty, and intrinsic supernova
magnitude and color dispersion needs to be explored.  The effects of
merging disparate supernova samples whose calibrations are not
perfectly correlated can be studied.  Equally important is an
exploration of systematic uncertainties incurred if incorrect models for
either dust-extinction or intrinsic supernova colors are used.
The generation of a realistic covariance matrix that faithfully
describes the calibration process would show where between
the ``Diagonal'' and ``Temperature'' models realistic observations will
lie.

For large data sets like those in~\cite{omnibus:2004} the new method presents
some practical numerical problems. In this study, a Fisher matrix
analysis has been performed, which provides reliable estimates
of the attainable precision. In the future, a Monte Carlo
approach should allow for a real fit to the data, as well as
facilitate the inclusion of non-Gaussian priors in, for instance, the
distributions of $A^i_V$ and $R^i_V$.
%
%
\section*{Acknowledgments} 
We wish to thank Eric Linder, Julian Borrill and Radek Stompor
for several useful discussions and Iwona Sakrejda at NERSC for her
help in getting our code to run on the PDSF system.
This work has been supported in part by the Director, Office of Science, Department of Energy under 
grant DE-AC02-05CH11231.  RM is partially supported by the National 
Science Foundation under agreement PHY-0355084. 
%
%

%
\end{document}